\documentclass[twocolumn,amsmath,amssymb,graphicx]{revtex4-1}
\usepackage{graphicx}
\usepackage{epstopdf}
\RequirePackage{fix-cm}

\begin{document}
\title{Particle-based simulations of steady-state mass transport at high P\'eclet numbers}
\author{Thomas M\"uller}
\author{Paolo Arosio}
\author{Luke Rajah}
\author{Samuel I.~A. Cohen}
\author{Emma V. Yates}
\author{Michele Vendruscolo}
\author{Christopher M. Dobson}
\author{Tuomas P.~J. Knowles}
\email{tpjk2@cam.ac.uk}

\affiliation{Department of Chemsitry, University of Cambridge, Lensfield Road, Cambridge CB21EW, United Kingdom}



\begin{abstract}
   Conventional approaches for simulating steady-state distributions of particles under diffusive and advective transport at high P\'eclet numbers involve solving the diffusion and advection equations in at least two dimensions. Here, we present an alternative computational strategy by combining a particle-based rather than a field-based approach with the initialisation of particles in proportion to their flux. This method allows accurate prediction of the steady state and is applicable even at high P\'eclet numbers where traditional particle-based Monte-Carlo methods starting from randomly initialised particle distributions fail. We demonstrate that generating a flux of particles according to a predetermined density and velocity distribution at a single fixed time and initial location allows for accurate simulation of mass transport under flow. Specifically, upon initialisation in proportion to their flux, these particles are propagated individually and detected by summing up their Monte-Carlo trajectories in predefined detection regions. We demonstrate quantitative agreement of the predicted concentration profiles with the results of experiments performed with fluorescent particles in microfluidic channels under continuous flow. This approach is computationally advantageous and readily allows non-trivial initial distributions to be considered. In particular, this method is highly suitable for simulating advective and diffusive transport in microfluidic devices.
\end{abstract}

\keywords{Steady-state mass transport, Convection, Diffusion, Microchannel}

\maketitle 


\section{Introduction}
\label{sec:intro}
Monitoring diffusion in microscopic channels offers a wide range of possibilities for 
determining the size of particles in solution or the viscosity of fluids 
\cite{Kamholz1999,Ismagilov2000,Costin2002,Costin2003,Squires2005}. For flow occurring at low Reynolds numbers, where viscous forces dominate over inertial ones, the velocity profile can be determined analytically for simple geometries (see for instance Ref.~\cite{Spiga1994}), a factor which gives a computationally inexpensive way of simulating advective transport. Exact results for advection coupled to diffusion are also known for certain geometries, notably in the case of Taylor dispersion \cite{Taylor1953}. In general, however, for predicting the steady-state distributions of particles in the presence of diffusion in arbitrary geometries, conventional strategies rely on solving the continuity equation in three dimensions \cite{Kamholz2001,Wu2005,Kennedy2012}.

Here we present a method for circumventing the requirement to work with the probability distribution directly, an approach similar to that used for field-flow fractionation \cite{Schure1988} and for obstacles in microchannels at low P\'eclet numbers \cite{Li2007}, and show that the downstream concentration profile of the analyte emerges from a particle \emph{flux}. Crucially, the computational cost of calculating this particle flux for discrete particles is only weakly dependent on the system geometry and the boundary or initial conditions. In particle-based simulations, the task of solving the Fokker-Planck equation \cite{Fokker1914,Smoluchowski1915,Planck1917,Risken1989}
\begin{equation}\label{eq:fp}
-\frac{\partial\rho(\vec{r},t)}{\partial t}=\left[\vec{\nabla}\cdot\vec{v}+D\vec{\nabla^2}\right]\rho(\vec{r},t)
\end{equation}
for the probability distribution $\rho(\vec{r},t)$ of the particles is replaced with that of solving the corresponding Langevin equation \cite{Langevin1908}
\begin{equation}\label{eq:langevin}
m\vec{\dot{v}}+\gamma \vec{v} = \vec{\Gamma}(t)
\end{equation}
for the motion of the particles with mass $m$. Here, $\vec{v}$ is the particle drift velocity, $D=kT/\gamma$ the diffusion coefficient with Boltzman constant $k$ and temperature $T$, $\gamma=6\pi\eta a$  the friction coefficient of a spherical particle with radius $a$, $\eta$ the fluid viscosity, and $\vec{\Gamma}(t)$ the random Langevin force field that ensures thermal equilibrium through $<\vec{\Gamma}(t)\cdot\vec{\Gamma}(t')>\ = 2kT\delta(t-t')$. From equation \ref{eq:langevin} it can be inferred that the mean-square displacement in each spacial dimension is $2Dt$.

Focusing on the Langevin equation allows a straightforward implementation of 
complex initial or boundary conditions as well as of force fields. Furthermore, 
this method is conceptually very simple as the stability analysis of the Langevin 
equation is much simpler than that of the Fokker-Planck equations that commonly 
take the form of complex partial differential equations which have to be iterated 
to convergence with specific boundary conditions in time and space.

The naive evaluation of the Langevin equation by establishing 
a steady-state flow of particles under continuous flow by creating new particles at every discrete time step is computationally very expensive as an increasingly large number of particles is required. A key realisation that makes this approach practically useful is that the steady state can in fact be simulated by loading all particles at $x=0$ and $t=0$ and propagating them 
until they have all left the area of interest, rather than introducing new 
particles at every step. In applications at low P\'eclet numbers, these particles can be loaded randomly \cite{Schure1988,Li2007}, since diffusion will randomise the distribution in any case. At high P\'eclet numbers, however, it is crucial how the particles are initiated. Here, we show that by initiating a (velocity-weighted) flux
\begin{equation}
\Phi(x=0,y,z)=C(x=0,y,z)\cdot v(y,z)/\bar{v},
\end{equation}
with $C(x=0,y,z)$ the initial concentration, $v(y,z)$ the fluid velocity at position $(y,z)$ and $\bar{v}$ the fluid velocity averaged over the channel cross-section, of a given distribution at time zero, we can simulate the steady-state downstream particle concentration by counting the number of time steps each Monte-Carlo trajectory spends in a predefined detection region. It is important to note that this method of initialisation and detection accounts for the velocity profile in microchannels. Furthermore, initialising the particle \emph{flux} instead of \emph{concentration} circumvents the detection bias that would occur otherwise for particles moving slowly through the detection region - for instance close to a channel wall. 

In order to demonstrate the applicability of our approach we have compared the predicted concentration profiles with experimentally obtained diffusive broadening of a stream of fluorescent colloids flowing continuously in a microfluidic channel at steady state, starting from their measured initial probability distribution. The readily achievable conditions for laminar flow on micrometer scales thereby enables the accurate calculation of the flow field inside the channel, thus leading to an accurate match between predicted and measured downstream particle distributions.

\section{Simulation Strategy}
\label{sec:methods}

\subsection{Description of the Problem}
\begin{figure*}
	\centering
	\includegraphics[width=0.5\textwidth]{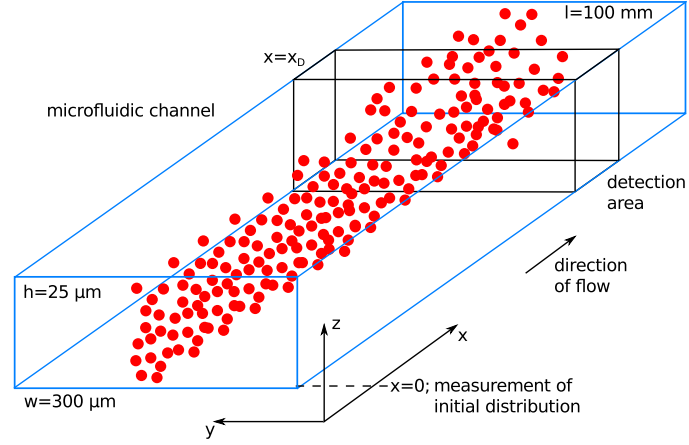}
	\caption{Schematic description of the problem addressed here. The geometry is defined by a microfluidic channel with a rectangular cross section and a width of $w=300~\mu\textrm{m}$, a height of $h=25~\mu\textrm{m}$, as well as a length of $l=100~\textrm{mm}$. A large number of monodisperse particles is loaded at $x=0$, according to a given distribution (which can be measured), and diffuses in all spatial dimensions under longitudinal laminar flow for a time determined by the distance from the starting point to the detection region.}
	\label{fig:ExpScheme}       
\end{figure*}

Figure~\ref{fig:ExpScheme} describes schematically the problem that we wish to solve. A flux of monodisperse particles is subjected to advective motion along a channel ($x$-axis) and diffusive motion in all directions at high lateral P\'eclet numbers - i.e., the time scale for advection along the channel is much shorter than the time scale for diffusion across the larger channel dimension. At a predefined distance from the initial point, the lateral distribution of the particles can be determined experimentally - e.g., by fluorescence microscopy - and predicted by simulation according to Eq.~\ref{eq:langevin}.

In our case, the channel dimensions are $w=300~\mu\textrm{m}$, $h=25~\mu\textrm{m}$, and $l=100~\textrm{mm}$ for width, height, and length, respectively. The flow rate is  $40~\mu\textrm{l/h}$, which leads to a Reynolds number of
\begin{equation}
Re=\frac{\rho v d_h}{\eta}\approx0.07\ll1,
\end{equation}
where $\rho=10^3~\textrm{kg/m}^3$ and $\eta=10^{-3}~\textrm{Pa}\cdot \textrm{s}$ are the fluid density and viscosity, $v\approx1.5~\textrm{mm/s}$ is the advective velocity along the channel, and $d_h=(2h\cdot w)/(h+w)=46~\mu\textrm{m}$ is the hydraulic diameter. For the P\'eclet number relating the time scales for diffusive and advective transport
\begin{equation}
Pe=\frac{\tau_D}{\tau_c}=\frac{\xi^2/2D}{l/v},
\end{equation}
we distinguish the cases along the height ($\xi=h$) and along the width ($\xi=w$) of the channel. Using a diffusion coefficient $D=kT/6\pi\eta r=8.6\times10^{-12}~\textrm{m}^2/\textrm{s}$ for a particle with $r=25~\textrm{nm}$ this yields $Pe=0.55$ in the vertical direction and $Pe=79$ in the horizontal direction. Note that these values mean that, due to the channel aspect ratio, during the time scale of the advective transport the particles sample the entire height but do not diffuse completely across the channel.

\subsection{Numerical Simulations}

In principle, a steady-state distribution of particles can be simulated by creating a new flux of particles at every discrete time step $\Delta t$ and propagating all of these sets. This approach, however, requires keeping track of a growing number of particles which is computationally extremely expensive and memory-intensive if a large number of propagation steps is desired. Furthermore, at low P\'eclet numbers, the particles can be initialised in a random distribution since their position will be randomised by diffusive mixing throughout the simulation. This, however, is not applicable at high P\'eclet numbers, and merely loading an initial concentration profile will fail since that does not take into account the velocity-weight of the particle motion across the channel cross section. Instead, we propose initiating a large flux $\Phi$ ($N>10^6$ particles) at $x=0$ and $t=0$, whereupon these particles are individually subjected to advective flow and diffusion, which enables determination of the steady state simply by counting the number of time steps $t_n$ the $k$-th particle spends in each lateral bin $Y_l$ in a predefined detection region $x_D$. This yields the concentration per bin via
\begin{equation}
C(Y_l)=\Sigma_k\Sigma_n \Delta_{[y_k(t_n) \in Y_l]}\cdot\Delta_{[x_k(t_n) \in x_D]},
\end{equation}
where $\Delta_{[y_k(t_n) \in Y_l]}$ is the Kronecker delta function, equal to 1 if $y_k(t_n)$ is within the region $Y_l$ and 0 otherwise.

\begin{figure*}
	\centering
	\includegraphics[width=0.7\textwidth]{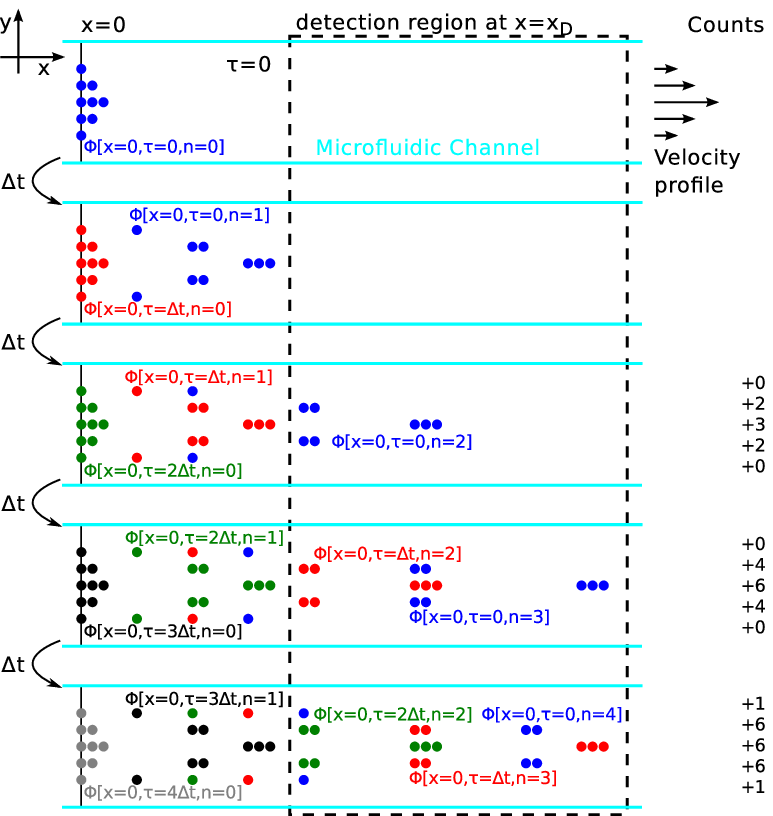}
	\caption{Schematic description of a traditional algorithm to determine the steady-state distribution of particles under flow, but utilising loading of a flux at a uniform distribution. At every time point $\Delta t$ a flux of particles (marked with different colours) is created at $x=0$, which is then propagated through the system. The final distribution is assessed by counting the particles resident in a detection area at $x=x_D$ until a steady state has been reached. In this notation, $\Phi$ denotes the flux of particles introduced at position $x=0$ and time $\tau$, propagated $n$ times (not to scale).}
	\label{fig:AlgScheme}
\end{figure*}

Figure \ref{fig:AlgScheme} illustrates the creation and propagation of fluxes at discrete time steps under advective flow. Specifically, at every time point $\tau$, a new flux of particles (symbolised by a different colour) is introduced at $x=0$ and the existing particles migrate along a channel according to a given velocity profile. For simplicity, diffusion is not considered in this schematic. From this graph it is clear that the flux $\Phi$ of particles depends only on the number of iterations enacted and not on the time step at which it was created. More formally,
\begin{equation}
\Phi[x=0,\tau=j\cdot\Delta t,n=l]=\Phi[x=0,\tau=0,n=l],
\end{equation}
where $\tau$ denotes the time point at which the particles were introduced into the device, and $n$ is the number of iterations effected on them. The $(x,y,z)$-dependence of $\Phi$ in this term is omitted for improved readability. Note that in practice the width of the detection region should be much larger than the distance propagated per step but much smaller than the distance from the start point to the detection region. Other than the effect of averaging the concentration profile over the length of the read area, the results of these simulations are not sensitive to the size of the detection region.

Therefore, the steady-state flux, which is given by
\begin{equation}
\Phi_l(x,y,z)=\sum_{j=0}^l\Phi[x=0,\tau=j\cdot\Delta t,n=l-j]
\end{equation}
with $l$ sufficiently large, can be written as
\begin{equation}
\Phi_l(x,y,z)=\sum_{j=0}^l\Phi[x=0,\tau=0,n=l-j].
\end{equation}
From there, the concentration in a detection region $x_D$ can be computed by
\begin{equation}
C(x,y,z)=\frac{1}{m}\sum_{r=l}^{l+m}\Phi_r(x,y,z)\Delta_{\left[x\in x_D\right]},
\end{equation}
where $m\cdot\Delta t$ corresponds to an exposure/integration time. This procedure results in 
\begin{equation}
C(x,y,z)=\frac{1}{m}\sum_{r=l}^{l+m}\sum_{j=0}^r\Phi[x=0,\tau=0,n=r-j]\Delta_{\left[x\in x_D\right]},
\end{equation}
which can now be simplified to
\begin{equation}\label{eq:concentration}
C(x,y,z)=\sum_{n=0}^s\Phi[x=0,\tau=0,n]\Delta_{\left[x\in x_D\right]},
\end{equation}
where $s$ equals the number of iteration steps needed for all the initialised particles to reach and exit the detection region. Equation~\ref{eq:concentration} is straightforwardly implemented by performing propagation steps individually on a large number of particles, $N$, until $x_k>x_D$ for all $k\in\left[0,N\right]$; note that $s$ may differ for each particle. Specifically, numerical assessment of the concentration is effected by:
\begin{enumerate}
	\item[(i)]{Initialising $N$ particles at $x=0$ in a given distribution multiplied by the longitudinal velocity $v_x(y_k,z_k)$ to obtain a flux $\Phi[x=0,\tau=0,n=0]$,}
	\item[(ii)]{Propagating each individual particle until $x_k>x_D$,}
	\item[(iii)]{Recording the $y$-values of each particles while $x_k\in x_D$.}
\end{enumerate}
This procedure results in a histogram of $y$ values that corresponds to the concentration distribution inside the channel. By also acquiring the $z$-values in $x_D$, the vertical profile can be simulated as well and compared, for example, with a confocal image \cite{Ismagilov2000}.

Taking into account diffusion and advection, propagation of the $k$th particle in all three spatial dimensions in a random walk is given by the discrete version of the integrated Langevin equation (\ref{eq:langevin})
	\begin{align}
	x_k^{(i+1)} &= x_k^{(i)} + v_x(y_k^{(i)},z_k^{(i)})\cdot \Delta t \nonumber \\
	& \quad  + \sqrt{2 D\Delta t} \cdot \textrm{Random}\{-1,+1\}\\
	y_k^{(i+1)} &= y_k^{(i)} + \sqrt{2 D \Delta t} \cdot \textrm{Random}\{-1,+1\}\\
	z_k^{(i+1)} &= z_k^{(i)} + \sqrt{2 D \Delta t} \cdot \textrm{Random}\{-1,+1\},
	\end{align}
with $v_x$ the advective flow velocity in $x$-direction, $\Delta t=0.3~\textrm{ms}$ the time interval of the simulations, $D$ the diffusion coefficient of the analyte in question, and Random selecting either -1 or +1. Note that this formulation of particle Brownian motion fulfils the fluctuation-dissipation theorem by construction \cite{Bedeaux1974}. Reflective boundary conditions are implemented at the channel walls. Additionally, any other form of concentration-independent migration such as electrophoresis can be incorporated readily into these steps. 

In the present case of laminar Poiseuille flow through a rectangular channel with no-slip boundary conditions, the velocity can be expressed analytically (see Ref.~\cite{Spiga1994}, for instance) by
\begin{equation}
v_x(y,z)=\alpha\sum_{i,j\textrm{ odd}} \frac{\sin\left(j\pi\frac{z}{h}\right)\sin\left(i\pi\frac{y}{w}\right)}{ij(i^2/w^2+j^2/h^2)} ,
\end{equation}
where $w$ and $h$ are the channel width and height, respectively, and $\alpha$ is a parameter such that integration over the channel cross section yields the predefined flow rate.

Depending on the number of particles, the length of the channel, the flow velocity, and the time step length $\Delta t$, a simulation takes between less than one minute up to a few hours. We note that this strategy does not take into account changes in viscosity due to changing particle concentrations - a limitation that is rarely a problem in practise in light of the low analyte concentration requirements afforded by present-day fluorescence microscopy techniques.

\subsection{Particle Initialisation}

\begin{figure*}
	\centering
	\includegraphics[width=0.5\textwidth]{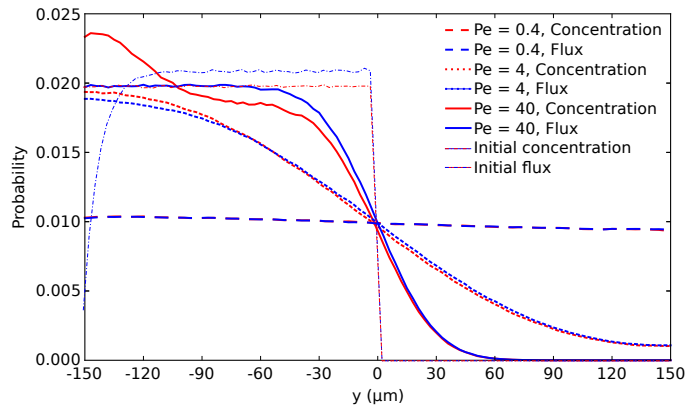}
	\caption{Comparison between simulations with particles initialised as a concentration (red lines) and as a flux (blue lines) for the case of low P\'eclet numbers ($Pe=0.4$, dashed lines), intermediate P\'eclet numbers ($Pe=4$, dotted lines), and high P\'eclet numbers ($Pe=40$, solid lines). Shown is the normalised concentration in the detection region at $l=50~\textrm{mm}$ determined via counting the number of time steps each Monte-Carlo trajectory spends in each bin along the channel width. The initialised relative concentration and flux are marked by the dash-dotted red and blue lines.}
	\label{fig:Loading}
\end{figure*}

The effect of the exact method of initialising the particles is demonstrated in Fig.~\ref{fig:Loading}. Simulations for $5\times10^6$ particles in a channel of width $w=300~\mu\textrm{m}$, height $h=25~\mu\textrm{m}$ and length $l=50~\textrm{mm}$ are performed at P\'eclet numbers of $Pe=0.4$, 4, and 40, by varying the flow rate from $Q=0.4~\mu\textrm{l/h}$ to $4~\mu\textrm{l/h}$ and $40~\mu\textrm{l/h}$. Here, the time step width was chosen to be 5~ms for $Pe=40$ and 25~ms for $Pe=4$ and $Pe=0.4$.

Particles were initialised uniformly in the left half of the channel according to their concentration (red lines) and their flux (blue lines), and detected by counting their occurrence in each of the 100 lateral bins in the detection region at $x_D=50~\textrm{mm}$.

In practise, initialisation according to concentration was achieved by placing each particle $k$ randomly along the height $h$ and half the width $w$ of the channel. Specifically,
\begin{eqnarray}
y_k & = & (0.5\cdot\textrm{Random}(0,1)-0.5)\cdot w,\\
z_k & = & (\textrm{Random}(0,1)-0.5)\cdot h,
\end{eqnarray}
where $\textrm{Random}(0,1)$ draws a uniformly distributed random number in the range between 0 and 1. Initialisation according to flux additionally compares a freshly drawn random number $p$ to the relative flow rate $v(y_k,z_k)/v_\textrm{max}$ and only places the particles if $p<v(y_k,z_k)/v_\textrm{max}$. This process is repeated until all $N$ particles could be placed. The detection algorithm increments a detector bin value $C(Y_l)$ for every time step a particle $k$ spends inside the detection region, i.e.
\begin{itemize}
	\item $C(Y_l)=0$ for all $l$;
	\item if $y_k(t_n) \in Y_l$ and $x_k(t_n) \in x_D$, the value $C(Y_l)$ is increased by 1;
	\item repeat until all particles $k$ have left the detection region $x_D$.
\end{itemize} 

The initial profiles are shown as the dash-dotted lines in Fig.~\ref{fig:Loading}. At low P\'eclet numbers, both the flux and concentration initialisation produce the correct profiles due to the almost complete randomisation of the particle positions by molecular diffusion. At intermediate, and even more so at high P\'eclet numbers, however, initialising a particle concentration leads to an overestimation of the relative intensity close to the edge of the channel as a consequence of the slower flow velocity yielding increased detection counts. 

These simulations demonstrate the necessity of considering an initial distribution that is proportional to a particle flux rather than a concentration. It should also be noted that the opposite method of initialising a concentration followed by velocity-weighting the detector counts will not lead to correct results. This is illustrated by the nearly perfect match of the readout profiles at low P\'eclet numbers.

\section{Experimental Validation of the Simulation Strategy}
\subsection{Materials and Methods}
In order to test the quantitative nature of the solution obtained from our 
simulations, we compared the results of this numerical simulation with experimental measurements 
of the transport of colloidal particles in microchannels. Microfluidic channels 
were fabricated at a width of $w=300~\mu\textrm{m}$, a height of 
$h=25~\mu\textrm{m}$ and a length of $l=100$~mm using standard soft 
lithography~\cite{Duffy1998}. Carbon nanopowder (633100, Sigma Aldrich, St.Louis MO, USA) was added to the elastomer prior to curing to create black devices that decrease the background from ambient and scattered light.

A suspension of fluorescent polystyrene colloids (FluoroMax G50, Thermo Scientific, Waltham MA, USA; nominal particle radius 25~nm with a uniformity better than $15\%$;  density $1.05~\textrm{g/cm}^3$) at a volume fraction of around $0.05\%$ and an auxiliary fluid, here deionised water, were supplied through a pipette tip inserted into the respective inlet. The analyte and auxiliary fluid were driven through the channel at a flow rate of $40~\mu\textrm{l/h}$ by a negative pressure on the device outlet, applied by a syringe pump (neMESYS, Cetoni GmbH, Korbu{\ss}en, Germany). In a nozzle area, the colloid suspension is focused in a narrow beam in between two flanking streams of auxiliary fluid. The flow of the particle suspension and of the auxiliary fluid makes up 10 and $90\%$ of the total flow, respectively, and is determined by the hydrodynamic resistance of the inlet channels.

To ensure full stabilisation of the Poiseuille flow profile through diffusion of 
vorticity, we allowed the analyte to migrate longitudinally over a distance of 8.6~mm 
before recording what we define as the ``initial'' distribution. Diffusive broadening 
was then measured at positions downstream of this point by 10, 20, 50, and 80 mm, corresponding to average migration times of 7, 14, 34, and 54~s.

All diffusion profiles were recorded through standard fluorescence microscopy using a cooled CCD camera (Evolve 512, Photometrics, Tucson AZ, USA) with a pixel size of 16x16~$\mu\textrm{m}^2$. We note that this detection method does not cover the $z$-dependence of the distribution resulting from the slower advective velocity closer to the channel walls \cite{Ismagilov2000,Kamholz2001}. However, since our experiments take place in the regime of small P\'eclet numbers in the vertical direction (diffusive migration distances are larger than the channel height) it is safe to assume that the analyte will sample the vertical velocity profile well enough to average out such effects.

\subsection{Comparison of Simulations and Measurements}
\label{sec:results}
\begin{figure*}
	\centering
	\includegraphics[width=0.7\textwidth]{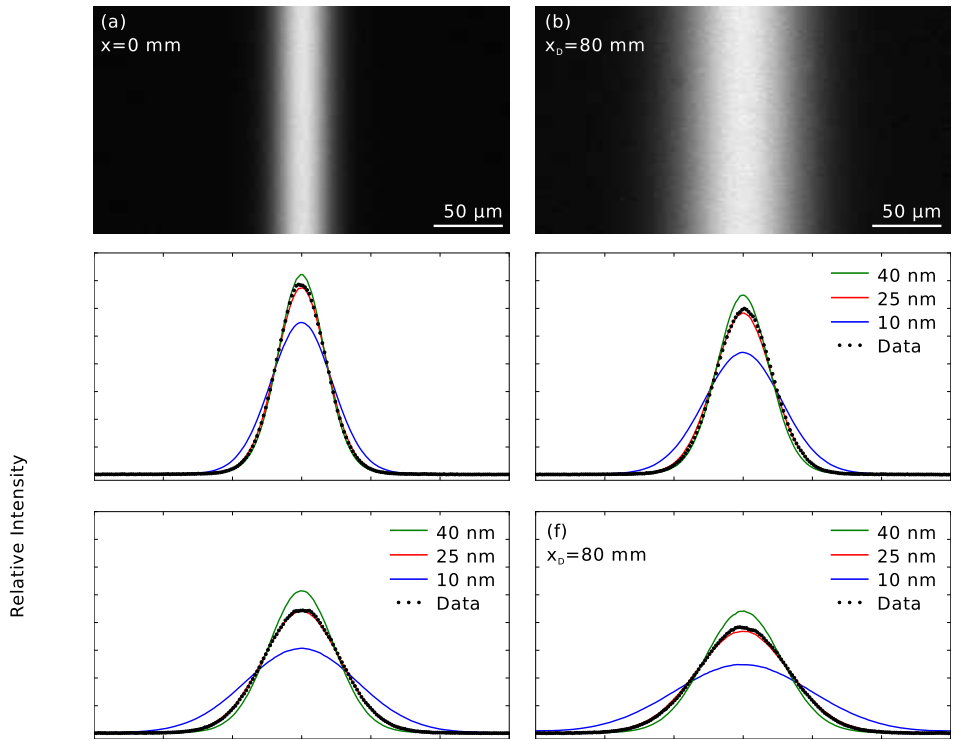}
	\caption{(a) and (b) Fluorescence microscopy images of a stream of fluorescent polystyrene colloids with a nominal radius of 25~nm at the starting point and after propagation over a distance of 80~mm, respectively. (c)-(f) Comparison between simulated and measured distributions at different downstream positions in the microfluidic channel. The initial experimental distribution is measured and used as the starting flux in the simulation. Of the simulated radii, the data sheet value of the colloids of $r=25$~nm clearly matches the experimental diffusion coefficient most closely, confirming the validity of this approach.}
	\label{fig:Data}
\end{figure*}

Figure \ref{fig:Data}(a) shows a fluorescence microscopy image of the initial distribution of fluorescent colloids at $x=0$. After migrating longitudinally over a distance of 80~mm, the beam has widened considerably in the lateral direction, as seen in Fig.~\ref{fig:Data}(b).

From fluorescence microscopy images at positions of 10, 20, 50, and 80~mm downstream from the initial point, the distributions of the analyte were extracted and are presented as the dotted lines in Fig.~\ref{fig:Data}(c)-(f). The simulated profiles for different, uniform particle sizes are shown as blue ($r=10~\textrm{nm}$), red ($r=25~\textrm{nm}$), and green lines ($r=40~\textrm{nm}$).

The excellent agreement between the measured distributions and the simulation using the data-sheet value of the colloids $r=25~\textrm{nm}$ demonstrates the feasibility of using arbitrary (but measured) initial probability distributions as starting points to simulate steady-state mass transport and verifies the accuracy of our approach. We emphasise that the simulated curves are not fits to the data, but are generated from first principles (using the measured initial profile) and the only free parameter is the particle size.

Nevertheless, we can quantify the deviations of the simulations from the measurement in a way analogous to the reduced $\chi^2$ value: calculating the sum of square errors per data point and normalising against an estimate of the experimental (via the standard deviations of the first and last 15 data points in each plot in Fig.~\ref{fig:Data}) we obtain
\begin{equation}
\frac{\Sigma_i(x^{(i)}_\mathrm{data}-x^{(i)}_\mathrm{sim})^2/N}{\sigma^2_\mathrm{exp}}=160,~380,~50,~310
\end{equation}
for the simulations with a radius of 25~nm shown in panels (c), (d), (e), and (f), respectively. In comparison, the corresponding numbers for radii of 10 or 40~nm are of the order of a thousand to tens of thousands. While the above values for $r=25~\textrm{nm}$ are clearly larger than one, our estimate for the variance of the experimental data only takes into account the camera read-out noise at low signal strengths and no deviations due to imperfections in the channels or from additional scatterers. In this sense, these values for the normalised square errors are to be understood as an upper bound. As such, these results show that the computed mass distributions approximate the measurements to better than a factor of $\sqrt{380}\approx20$ of the noise level.

\section{Conclusion}

We have proposed and validated experimentally a strategy for simulating steady-state mass transport of particles under flow that relies on determining the trajectories of individual particles initialised according to their generating flux rather than solving the Fokker-Planck equations of the probability distributions. This approach allows for studies of systems at high P\'eclet numbers and provides a straightforward way of incorporating measured initial profiles as well as facile implementation of force fields, such as electric or gravitational forces. Microfluidics as well as nanofluidics provide particularly well-suited application platforms for this approach since devices are typically operated in the regime of Poiseuille flow and diffusion is widely used for controlled mixing, separating and analysing particles in solutions.

\begin{acknowledgments}
	Financial support from the Biotechnology and Biological Sciences Research Council (BBSRC), the European Research Council (ERC), the Frances and Augustus Newman Foundation as well as the Swiss National Science Foundation is gratefully acknowledged.
\end{acknowledgments}



\end{document}